\documentclass[11pt]{article}
\usepackage{moriond,epsfig}

\def\be{\begin{equation}}
\def\ee{\end{equation}}
\def\bea{\begin{eqnarray}}
\def\eea{\end{eqnarray}}

\begin{document}
\vspace*{4cm}
\title{A Chameleon Primer}

\author{ Ph.~Brax, C.~van de Bruck and A.-C.~Davis }

\address{  Service de Physique Th\'eorique, CEA/DSM/SPhT,
Unit\'e de recherche associ\'ee au CNRS, CEA--Saclay F--91191
Gif/Yvette cedex, France\\ 
  Department of Applied Mathematics,
University of Sheffield, Hounsfield Road, Sheffield S3 7RH, United
Kingdom\\ 
  Department of Applied Mathematics and Theoretical
Physics, University of Cambridge, Clarkson Road, UK\\}

\maketitle\abstracts{We review some of the properties of chameleon
theories. Chameleon fields are gravitationally coupled to matter
and evade gravitational tests thanks to two fundamental
properties. The first one is the density dependence of the
chameleon mass. In most cases, in a dense environment, chameleons
are massive enough to induce a short ranged fifth force. In other
cases, non-linear effects imply the existence of a thin shell
effect shielding compact bodies from each other and leading to an
irrelevant fifth force. We also mention how a natural extension of
chameleon theories can play a role to solve the PVLAS versus CAST
discrepancy.}

\section{Introduction}
Fifth force experiments such as the Cassini satellite experiment
put stringent bounds on the gravitational coupling of nearly
massless scalar particles. Future satellite tests of  fifth forces
and putative violations of the equivalence principle will even
lead to stronger constraints. As no scalar field has ever been
observed, these bounds would not be so dramatic if the existence
of nearly massless fields was not suggested by the late time
acceleration of the universe expansion\cite{supernovae1,supernovae2}. In fact, these constraints
have fundamental consequences for models of dark energy. Indeed,
models of dark energy known as quintessence\cite{wetterich,ratra,stein,quintreview}
require the existence of a runaway scalar field with a tiny mass now,
of order $H_0\sim 10^{-43}$ GeV. The range of the interactions mediated
by the quintessence scalar field is of order of the Hubble horizon size.

Hence, unless the quintessence field has a very small
gravitational interaction with ordinary matter, fifth force
experiments are not compatible with a quintessence scenario. For
instance, embedding quintessence models in spontaneously broken
supergravity proves to be extremely difficult as the gravitational
couplings are generically large\cite{braxmartin}. Within string
theory, the dilaton has been argued to be a quintessence candidate
provided the coupling to matter is universal and possesses a
minimum playing the role of an attractor where all gravitational
problems are evaded\cite{damour}. Of course, it would be extremely
interesting to confirm this possibility explicitly. String moduli
fields are also natural candidates for quintessence.
Unfortunately, their gravitational coupling is generically of
order one. On the other hand, there exists a well-motivated scalar
field with a small gravitational field: the radion measuring the
inter-brane distance in Randall-Sundrum scenarios. In this case,
the gravitational coupling of the radion to matter on a warped
brane is suppressed by the warp factor and becomes very small for
a large radion.

Chameleon field theory combine both a quintessence-like behaviour
leading to dark energy at late time and a gravitational coupling
to matter which can be large\cite{cham1,cham2}.
So how come they are not definitely
ruled out by fifth force experiments? In fact, it is useful to
draw an analogy with photons. In some circumstances, photons do
get a mass which alters their properties. This is notoriously the
case in superconductors where the Meissner effect (the fact that
the magnetic field is expelled from a superconductor) can be seen
as the result of the Higgs mechanism with a mass given to the
photons\cite{weinberg2}. In less extreme situations, like in a crystal, photons
are slowed down when interacting with matter. Similar phenomena
can occur for scalar fields. Typically, scalar particles have an
effective potential obtained as a combination of the bare
potential appearing in the Lagrangian and a term proportional to
the matter density. This effective potential may have a density
dependent minimum. In this case, we call the field a chameleon as
its mass depends on the environment.

Chameleon fields are generically more massive in a dense
environment. This is enough to evade the gravitational bounds in
most cases. Indeed, the range of the chameleon mediated force
becomes too small to be detected. Even when this is not the case,
chameleon theories may enjoy another non-trivial property: the
existence of thin shells. More precisely, the field created by a
massive object may be essentially trapped inside the massive body.
In this case, the interaction between massive bodies is
essentially non-existent. Combining these two effects, one can
build satisfying examples of chameleon theories. We will review
their main properties here.

Recently, the PVLAS experiment has measured the dichroism of light
propagating through a magnetic field\cite{PVLAS}. This can be understood by
coupling a scalar field to  photons. In this case, one can use the
environment dependent mass of chameleon fields to generate a large
mass for the chameleon in the sun. Therefore, chameleons would not
be produced by the Primakov effect and therefore the CAST
experiment\cite{CAST} would not see the photon regeneration by inverse
Primakov effect. We will present some of these ideas very briefly.

\section{Scalar-Tensor Theories}

\subsection{Coupling to matter}
Chameleon fields appear in scalar--tensor theories of gravity\cite{scalartensor}. We
start with a discussion of these theories.
We consider theories where a scalar field $\phi$ couples both to
gravity and matter, generating a potential fifth force. The
Lagrangian of such scalar--tensor theories reads
\begin{equation}
S = \frac{1}{2\kappa_4^2}\int d^4 x\sqrt{-g} (R- (\partial \phi)^2
-2\kappa_4^2 V(\phi))
\end{equation}
 Matter couples to both gravity and the scalar field
according to
\begin{equation}
S_m(\psi, A^2(\phi)g_{\mu\nu}),
\end{equation}
where $\psi$ is a matter field and $A$ is an arbitrary function of
$\phi$. The Klein--Gordon equation can be written in terms of an
effective potential
\begin{equation}\label{Veff}
V_{\rm eff}(\phi)=V(\phi) +\rho_m  A(\phi).
\end{equation}
The effective potential depends on the environment through the
matter energy density $\rho_m$.  We will assume that $V(\phi)$ is
a runaway potential and for the models we consider $A(\phi)$
increases with $\phi$. In that case the potential has a minimum
whose location depends on $\rho_m$, i.e. on the environment. Such
a field has been called a chameleon field.

 The field $\phi$ acts
on all types of matter and, in the Einstein frame, there is a new
force associated with the scalar field
\begin{equation}\label{fifthforce}
F_{\phi} = - \kappa_4 m \alpha_\phi\frac{\partial \phi}{\partial
x_\mu},
\end{equation}
where $m$ is the mass of the test particle in the Einstein frame
and
\begin{equation}
\alpha_\phi = \frac{\partial \ln A}{\partial \kappa_4 \phi}
\end{equation}
 The force $F_{\phi}$ cannot be too large, otherwise
experiments would have already detected it.

For massless fields $V(\phi)\equiv 0$ and a point-like matter
source,  the Klein-Gordon equation becomes
\begin{equation}
\Delta \phi= -\kappa_4 m \alpha_\phi \vert_{r=0}\delta^{(3)}(r)
\end{equation}
where $m= A(\phi)\vert_{r=0} m_0$ is the Einstein frame mass and
$m_0$ the bare mass of the source. The resulting field $\phi=
-\kappa_4 \alpha_\phi\vert_{r=0} /4\pi r$ leads to a force between
bodies $F_\phi= 2G_N \alpha_{\phi\vert_{r=r_1}}
\alpha_{\phi\vert_{r=r_2}} m_1 m_2 /r_{12}$ where $\kappa_4^2=8\pi
G_N$. This produces a fifth force where
\begin{equation}
F_\phi= 2\alpha_1\alpha_2 F_{\rm Newton}
\end{equation}
and $\alpha_1=\alpha_{\phi\vert_{r=r_1}}$. The Cassini experiments
impose that $\alpha_\phi^2\le 5. 10^{-5}$ for a constant coupling.
Hence massless particles (or nearly massless particles with a mass
less than $10^{-3}$ eV) must have a very small coupling to
gravity. Chameleon field theories enable to overcome this
obstacle.

\subsection{The radion}

A simple and interesting example of non-trivial coupling to
gravity is provided by the Randall-Sundrum scenario where matter
is confined on 4d hyperplanes embedded in an $AdS_5$
vacuum\cite{randallsundrum}. The two boundaries of space-time are
called the UV and the IR brane reflecting the fact that the metric
is warped. Distances on the IR branes are warped down compared to
scales on the UV. Consider now matter on the UV brane of positive
tension. The coupling of matter to gravity depends on the radion
field $\phi$ (for a derivation of the following equations and
references, see e.g. the review\cite{branereview})
\begin{equation}
A(\phi)=\cosh \frac{\kappa_4 \phi}{\sqrt 6}
\end{equation}
where the inter-brane distance is
\begin{equation}
d=-l \ln (\tanh \frac{\kappa_4\phi}{\sqrt 6})
\end{equation}
For small distances compared to the AdS curvature $l$, the
coupling becomes
\begin{equation}
A(\phi)= \frac{1}{2} e^{\kappa_4 \phi/\sqrt 6}
\end{equation}
The gravitational coupling is constant
\begin{equation}
\alpha_\phi= \frac{1}{\sqrt 6}
\end{equation}
Of course, this is too big for the Cassini bound. However, in
this case and in the large interbrane-distance limit, the chameleon mechanism
can be applied to hide the interaction mediated by the radion\cite{chamrad}
by introducing a bare potential for the radion field.

\subsection{Chameleon Cosmology}

We  concentrate on a particular model where $A(\phi)=
e^{\beta\phi}$ and $\beta=O(1)$.
 We consider the family of potential
\begin{equation}
V=M^4 f((\frac{M}{ \phi})^n)
\end{equation}
where $f$ leads to ordinary quintessence with a long time tracking
solution. A typical example is provided by $f(x)=e^x$. As $\phi
\gg M$ now, the potential is nothing but
\begin{equation}
V=M^4 + \frac{M^{4+n}}{\phi^n}
\end{equation}
Cosmologically, it mimics a cosmological constant. For
gravitational tests, only the Ratra--Peebles part of the potential
matters.

This model satisfies the chameleon property of having a
$\rho$--dependent minimum. As $\beta=0(1)$, the coupling of matter
to the chameleon field is large and may be in conflict with
experiments. We will study the gravitational aspects in the next
section. Here we concentrate on cosmological properties.

In a Friedmann--Robertson--Walker Universe, the (non)-conservation
of matter equation reads
\begin{equation}
\dot \rho +3H \rho =\alpha_\phi  \dot \phi \rho.
\end{equation}
leading to
\begin{equation}
\rho=A(\phi) \rho_m,\  \rho_m=\frac{\rho_0}{a^{3(1+w_m)}}
\end{equation}
while the Klein--Gordon equation can be written in terms of an
effective potential
\begin{equation}\label{effpot}
V_{\rm eff}(\phi)=V(\phi) +\rho_m(1-3w_m)  A(\phi).
\end{equation}

Let us now go through the different cosmological eras\cite{cham2}. During
inflation the chameleon potential has an effective minimum which
is time-independent. Moreover, as the mass of the chameleon field
at the minimum is $m\gg H$, the field oscillates rapidly and
converges to the minimum extremely rapidly,  behaving like a dust
component. By the end of inflation, the field is stuck at the
minimum. As inflation stops and the radiation era starts, the
minimum is pushed far away (as it depends only of non-relativistic
matter). The field is therefore in an overshooting regime where it
becomes kinetically dominated, being far  away to the right of the
potential. The field overshoots before stopping at  $\phi_{stop}=
\phi_{in}+ \sqrt{6\Omega_\phi^i}m_{\rm Pl}$ where $\Omega^i_\phi$
is the initial chameleon fraction density. After stopping the
field is in an undershooting position. In that case, the field
would remain still until either being caught up by the minimum or
the beginning of the matter era. When caught by the minimum the
field oscillate and converges to the minimum, which is a tracker
solution. This follows from the fact that $m\gg H$ at the minimum
throughout the history of the Universe. The field converges to the
minimum  faster than $a^{-3}$ due to the time variation of the
mass at the minimum.

In fact if the field is far away from the minimum after
overshooting, it is sensitive to  short bursts when relativistic
particles become non--relativistic
\begin{equation}
\ddot\phi +3H\dot\phi =\frac{\beta}{m_{\rm Pl}}T^\mu_\mu
\end{equation}
as, during such periods, the trace of the energy momentum tensor
$T^\mu_\mu$  of the decoupling species is temporarily
non-vanishing, resulting in a kick\cite{dn} of order of a fraction of
$\beta$. Taking into account all these kicks, the field decreases
by about $\Delta\phi \sim -\beta m_{\rm Pl}$. By BBN, either the
field is close to the minimum, in which case the electron kick
which occurs during BBN does not lead to a large variation of
$\phi$ during BBN, or the field is still far away from the minimum
in which case the electron kick leads to large variations of
$\phi$ and therefore of masses
\begin{equation}
\vert \frac{\Delta m}{m}\vert =\beta \vert \frac{\Delta
\phi}{m_{\rm Pl}}\vert
\end{equation}
 the latter case being  excluded. As a result, the
initial value of $\phi$ cannot be larger that one  and
$\Omega_\phi^i\le 1/6$, a weaker bound than in quintessence. Once
at the minimum by BBN, the field follows the attractor in the
matter era. Once the vacuum energy dominates, the matter density
decreases extremely fast. The chameleon field follows the minimum
until $m\sim H$ where it starts lagging behind eventually having
the same evolution as a quintessence field with no coupling to
matter.

\section{Gravitational Tests}
\subsection{The massive chameleon}
The effective potential with $f(x)=e^x$ leads to a stabilisation
of the scalar field for
\begin{equation}\label{minimum}
\phi= \left(\frac{n \Lambda^{4+n}M}{ \rho}\right)^{1/(n+1)},
\end{equation}
where $\rho$ is the matter energy density . The mass at the bottom
of the potential is given by
\begin{equation}
m^2= n(n+1) \frac{\Lambda^{n+4}}{\phi^{n+2}}
\end{equation}
In the atmosphere, the mass of chameleons is larger than $10^{-3}$
eV implying no consequence for Galileo's Pisa experiment and
similar tests.

\subsection{The thin shell}
Let us now consider a situation where the gravitational
experiments are performed on a body embedded in a surrounding
medium. The body could be a small ball of metal in the atmosphere
or a planet in the inter-planetary vacuum. The effective
potential~(\ref{effpot}) is not the same inside the body and
outside because $\rho _{\rm m }$ is different. The effective
potential can be approximated by
\begin{equation}
\label{approxVeff}
 V_{\rm eff}\simeq \frac12 m_\phi^2(\phi-\phi_{\rm min})^2\, ,
\end{equation}
 As already
mentioned the minimum and the mass are different inside and
outside the body. We denote by $\phi_{\rm b}$ and $m_{\rm b}$ the
minimum and the mass in the body and by $\phi_{\infty}$ and
$m_{\infty}$ the minimum and the mass of the effective potential
outside the body. Then, the Klein-Gordon equation reads
\begin{equation}
\label{radialKG} \frac{{\rm d}^2\phi}{{\rm
d}r^2}+\frac{2}{r}\frac{{\rm d}\phi}{{\rm d}r}= \frac{\partial
V_{\rm eff}}{\partial \phi}\, ,
\end{equation}
where $r$ is a radial coordinate.  Requiring that $q$ remains
bounded inside and outside the body and joining the interior and
exterior solutions, one can determine the complete profile which
can be expressed as
\begin{eqnarray}
\phi_{<}\left(r\right) &=& \phi_{\rm b}+\frac{R_{\rm
b}\left(\phi_{\infty}-\phi_{\rm b}\right)\left(1+m_{\infty }R_{\rm
b}\right)}{\sinh \left(m_{\rm b}R_{\rm b}\right)\left[m_{\infty
}R_{\rm b}+m_{\rm b}R_{\rm b} \coth \left(m_{\rm b}R_{\rm
b}\right)\right]}\frac{\sinh \left(m_{\rm b}r\right)}{r}\, ,
\qquad r\le R_{\rm b}\, , \nonumber \\  \phi_{>}\left(r\right) &=&
\phi_{\infty}+R_{\rm b}\left(\phi_{\rm b}-\phi_{\infty}\right)
\frac{m_{\rm b}R_{\rm b}\coth\left(m_{\rm b}R_{\rm
b}\right)-1}{\left[m_{\infty}R_{\rm b}+m_{\rm b}R_{\rm
b}\coth\left(m_{\rm b}R_{\rm b}\right)\right]}\frac{{\rm
e}^{-m_{\infty}\left(r-R_{\rm b}\right)}}{r} \, , \qquad r\ge
R_{\rm b}\, \nonumber \\
\end{eqnarray}
Assuming, as it is always the case in practise, that $m_{\rm b}\gg
m_{\infty}$, $m_{\rm b}R_{\rm b}\gg 1$, one has
\begin{equation}
\frac{\partial \phi_>(r)}{\partial r}\simeq -\frac{R_{\rm b}}{r^2}
\left(\phi_{\infty}-\phi_{\rm b}\right)\, ,
\end{equation}
from which we deduce that the acceleration felt by a test particle
is given by
\begin{equation}
a=\frac{G_Nm_{\rm b}}{r^2}\left[1+\frac{\alpha
_\phi\left(\phi_{\infty}-\phi_{\rm b}\right)}{\Phi _{_{\rm
N}}}\right]\, ,
\end{equation}
where $\Phi _{_{\rm N}}=G_Nm_{\rm b}/R_{\rm b}$ is the Newtonian
potential at the surface of the body. Therefore, the theory is
compatible with gravity tests if
\begin{equation}
\frac{\alpha _\phi\left(\phi_{\infty}-\phi_{\rm b}\right)}{\Phi
_{_{\rm N}}}\ll 1\, .
\end{equation}
Large compact bodies have a thin shell implying that no distortion
of solar system  planetary orbits are predicted. Lunar ranging
experiments are not affected either.

\subsection{Chameleon in a cavity}
 Gravitational experiments on earth and future satellite
 experiments involve vacuum chambers which can be modelled out as
 spherical cavities of radius $R$. Solving the chameleon equations in this
 situation, following the same method as in the previous
 subsection, we find that the mass of the chameleon field inside
 the cavity is determined by the resonance equation
 \begin{equation}
 \frac{\sinh m_0 R}{m_0R} = n+2
 \end{equation}
 Having determined $m_0$, one can deduce the value of the field
 $\phi_0$ inside the cavity. Notice that for most values of $n$ we
 have
 \begin{equation}
 m_0 R= O(1)
 \end{equation}
 When $\beta= O(1)$, the mass of the chameleon in gravitational
 experiments on earth is of order $1/R$ and is too large to evade
 gravitational tests (the range is given by $R\sim 1 $ m).
 Fortunately, typical test bodies on earth have a thin shell
 implying no deviation from Newton's law for Eotvos or Eotwash experiments\cite{eric}. Future satellite
 experiments are such that test bodies do not have a thin shell.
 Hence large deviations from Newton's law are predicted. When
 $\beta \gg 1$, tests bodies have  a thin shell and satellite
 experiments would not see any deviation. (For a discussion
of the case $\beta \gg 1$, see reference\cite{shaw}).

\section{PVLAS vs CAST}

Recently, the coupling of a scalar field to photons have been
invoked in order to explain the PVLAS results on dichroism\cite{PVLAS}.
The scalar field is required to have a mass of order
$10^{-12}$ GeV and a coupling strength suppressed by a scale of
order $M=10^6$ GeV.  The coupling to photons is given by
\begin{equation}
-\frac{1}{4} \int d^4 x e^{\phi/M} F_{\mu\nu}F^{\mu\nu}
\end{equation}
 The results of the PVLAS collaboration are in conflict with
astrophysical bounds such as CAST\cite{CAST}, which for the same
mass for the scalar field, require much smaller couplings
($M>10^{10}$GeV).

The chameleon mechanism can help in explaining the PVLAS results
and, at the same time, be in agreement with astrophysical bounds\cite{PVLASus}.
 Our model  is of the
scalar-tensor type
\begin{eqnarray}
S&=&\int d^4x \sqrt{-g}\left(\frac{1}{2\kappa_4^2}R-
g^{\mu\nu}\partial_\mu\phi \partial_\nu \phi -V(\phi)
-\frac{e^{\phi/M}}{4} F^2\right)\nonumber \\ &+& S_m( e^{\phi/M}
g_{\mu\nu},\psi_m)
\end{eqnarray}
where $S_m$ is the matter action and the fields $\psi_m$ are the
matter fields.  The effective gravitational coupling is given by
\begin{equation}
\beta= \frac{m_{\rm Pl}}{M},
\end{equation}
and therefore very large ($\beta = 10^{13}$) when assuming the
results from the PLVAS experiment ($M=10^6$~GeV). To prevent large
deviations from Newton's law one must envisage non--linear effects
shielding massive bodies from the scalar field. One natural
possibility is that the scalar field $\phi$ coupled to photons has
a runaway (quintessence)--potential leading to the chameleon
effect. For exponential couplings, this is realised when
\begin{equation}\label{poti}
V(\phi)= \Lambda^4\exp (\Lambda^n/\phi^n) \approx \Lambda^4 +
\frac{\Lambda^{4+n}}{\phi^n}
\end{equation}
In the presence of matter, the dynamics of the scalar field is
determined by an effective potential
\begin{equation}\label{effpot1}
V_{\rm eff }(\phi)=\Lambda^4\exp (\Lambda^n/\phi^n)+
e^{\phi/M}(\rho +\frac{{\bf B}^2}{2})
\end{equation}
where $\rho$ is the energy density of non-relativistic matter.

As already mentioned, the PVLAS experiment is in conflict with the
CAST experiment on the detection of scalar particles emanating
from the sun, as it requires $M\ge 10^{10}$ GeV. However, this
bound does not apply when the mass of the scalar field in the sun
exceeds $10^{-5} \rm GeV$. Let us evaluate the mass of the
chameleon field inside the sun. Furthermore, from the effective
potential one obtains
\begin{equation}\label{relati}
m_{\rm sun}= m_{\rm lab} \left(\frac{\rho_{\rm sun}}{\rho_{\rm
lab}}\right)^{(n+2)/2(n+1)}.
\end{equation}
Now $\rho_{\rm sun}/\rho_{\rm lab}\approx 10^{14}$ and, with
$n=0(1)$, one finds
\begin{equation}
m_{\rm sun} \sim 10^{-2} {\rm GeV} \gg 10^{-5} {\rm GeV}
\end{equation}
implying no production of chameleons in the sun. Hence, the CAST
experiment is in agreement with the chameleon model due to the
fact that the chameleon field is very massive in the sun.

\section{Conclusion}

We have given an brief overview of chameleon field theories. They
provide exciting new mechanisms for both gravitation and cosmology.

A scalar field coupled to matter can be problematic, since it
mediates a new force. But if the field self-interacts in a non-linear way,
as it is the case in chameleon field theories, the effect of the
field can be hidden from current experiments. As we pointed out,
future experiments will be able to search for such chameleon fields.
We have speculated that the PVLAS anomaly finds a natural interpretation
within these theories.

\section*{Acknowledgments}
We would like to thank out friends and collaborators on various
aspects of chameleon theories:  A. Green, J. Khoury, D. Mota, D.
Shaw and A. Weltman.


\section*{References}
\begin{thebibliography}{99}
\bibitem{supernovae1} S.~Perlmutter et al [Supernovae Cosmology Project],
Astrophys.J. {\bf 517}, 565 (1999)

\bibitem{supernovae2} A.Riess et a. [Supernovae Search Team],
Astron. J. {\bf 116}, 1009 (1998)

\bibitem{wetterich} C. Wetterich, Nucl. Phys. B{\bf 302}, 668 (1988)

\bibitem{ratra} B. Ratra and P. Peebles, Phys.Rev.D {\bf 37}, 3406 (1988)

\bibitem{stein} R.R. Caldwell, R. Dave and P. Steinhardt, Phys.Rev.Lett. {\bf 80}, 1582 (1998)

\bibitem{quintreview} for a review, see E.J. Copeland, M. Sami and S. Tsujikawa,
Int.~J.~Mod.~Phys.D {\bf 15}, 1753 (2006)

\bibitem{braxmartin} Ph. Brax and J. Martin, JCAP{\bf 0611}, 008 (2006)

\bibitem{damour} T.~Damour, F.~Piazza and G. Veneziano, Phys.Rev.D~{\bf 66},046007 (2002) 

\bibitem{cham1} J. Khoury and A. Weltman, Phys.Rev.D {\bf 69}, 044026 (2004)

\bibitem{cham2} Ph. Brax, C. van de Bruck, A.-C. Davis,
J. Khoury and A. Weltman, Phys.Rev.D {\bf 70}, 123518 (2004)

\bibitem{weinberg2} S. Weinberg, The Quantum Theory of Fields - Volume 2, Cambridge University 
Press (1996)

\bibitem{PVLAS} E.~Zavatini, et al [PVLAS collaboration],
Phys.Rev.Lett. {\bf 96}, 110406 (2006)

\bibitem{CAST} Zioutas, K. et al [CAST collaboration], Phys.Rev.Lett. {\bf 94}, 121301 (2005)

\bibitem{scalartensor} see e.g. Y. Fujii and K-I. Maeda, {\it The Scalar-Tensor Theory of Gravitation},
Cambridge University Press (2003)

\bibitem{randallsundrum} L. Randall and R. Sundrum,
Phys.Rev.Lett {\bf 83}, 4690 (1999)

\bibitem{branereview} Ph. Brax, C. van de Bruck and A.-C. Davis,
Rept.Prog.Phys. {\bf 67}, 2183 (2004)

\bibitem{chamrad} Ph. Brax, C. van de Bruck and A.-C. Davis,
JCAP {\bf 0411}, 004 (2004)

\bibitem{dn} T. Damour and K. Nordtvedt, Phys.Rev.D {\bf 48}, 3436 (1993)

\bibitem{eric} C.D. Hoyle, B.R. Heckel, E.G. Adelberger, J.H. Gundlach, D.J. Kapner 
and H.E. Swanson, Phys.Rev.Lett. {\bf 86}, 1418 (2001)

\bibitem{shaw} D.F.~Mota and D.J. Shaw, Phys.Rev.D {\bf 75}, 063501 (2007)

\bibitem{PVLASus} Ph.Brax, C. van de Bruck and A.-C. Davis, hep-ph/0703243

\end{thebibliography}
\end{document}